\documentstyle[11pt]{article}
\let\la=\label

\newfont{\bbbold}{msbm10 scaled \magstep1}

\newfont{\goth}{eufm10 scaled \magstep1}

\def\a{\alpha}
\def\b{\beta}
\def\c{\gamma}
\def\d{\delta}
\def\e{\epsilon}\def\vare{\varepsilon}

\def\l{\lambda}\def\L{\Lambda}
\def\m{\mu}
\def\n{\nu}

\def\r{\rho}
\def\s{\sigma}
\def\t{\tau}
\def\th{\theta}

\def\beq{\begin{equation}}\def\eeq{\end{equation}}
\def\beqa{\begin{eqnarray}}\def\eeqa{\end{eqnarray}}
\def\barr{\begin{array}}\def\earr{\end{array}}

\def\del{\partial}
\def\lb{{\bar\lambda}}
\def\cb{{\bar\chi}}

\def\cH{\cal H}
\def\nab{\nabla}
\newcommand{\w}[1]{\\[0.#1cm]}

\def\pl#1#2#3{Phys.~Lett.~{\bf {#1}B} (19{#2}) #3}

\let\bm=\bibitem

\def\nn{\nonumber}
\def\bd{\begin{document}}
\def\ed{\end{document}}
\def\ba{\begin{array}}
\def\ea{\end{array}}
\def\bea{\begin{eqnarray}}
\def\eea{\end{eqnarray}}
\def\ft#1#2{{\textstyle{{\scriptstyle #1} 
\over {\scriptstyle #2}}}}
\def\fft#1#2{{#1 \over #2}}
\def\sst#1{{\scriptscriptstyle #1}}
\def\oneone{\rlap 1\mkern4mu{\rm l}}
\newcommand{\be}{\begin{equation}}
\newcommand{\ee}{\end{equation}}
\newcommand{\eq}[1]{(\ref{#1})}
\def\eqs#1#2{(\ref{#1}-\ref{#2})}
\def\det{{\rm det\,}}
\def\tr{{\rm tr}}
\newcommand{\ho}[1]{$\, ^{#1}$}
\newcommand{\hoch}[1]{$\, ^{#1}$}
\newcommand{\tamphys}{\it\small Center for Theoretical Physics, 
Texas A\&M University, College Station, TX 77843, USA}
\newcommand{\kings}{\it\small Department of Mathematics, King's College,
London, UK}

\begin{document}
\pagestyle{empty}
\rightline{King's College/KCL-TH-98-11}
\rightline{CTP TAMU-18/98}
\rightline{hep-th/9806050}
\vspace{1.5truecm}
\centerline{\Large \bf Anomaly-Free Tensor-Yang-Mills System}
\vspace{0.5cm}
\centerline{ \Large\bf and Its Dual Formulation}
\vspace{1truecm}
\centerline{\bf P.S. Howe}
\medskip
\centerline{Department of Mathematics}
\centerline{King's College, London, UK}
\vspace{.5truecm}
\centerline{and}
\vspace{.5truecm}
\centerline{\bf E. Sezgin\footnote{
Supported in part by the U.S. National Science Foundation, under
grant PHY--9722090}}
\medskip
\centerline{Center for Theoretical Physics, Texas A\&M University,}
\centerline{College Station, Texas 77843, USA}
\vspace{1.3truecm}
\centerline{ABSTRACT}
\vspace{.5truecm}

We consider the $(1,0)$ supersymmetric Yang-Mills multiplet coupled to a
self-dual tensor multiplet in six dimensions. It is shown that the
counterterm required to cancel the one-loop gauge anomaly modifies the
classical equations of motion previously obtained by Bergshoeff, Sezgin
and Sokatchev (BSS). We discuss the supermultiplet structure of the
anomalies exhibited in the resulting equations of motion. The anomaly
corrected field equations agree with the global limit, recently obtained
by Duff, Liu, Lu and Pope, of a matter coupled supergravity theory in
six dimensions. We also obtain the dual formulation of the BSS model in
which the tensor multiplet is free while the field equations of the
Yang-Mills multiplet contain the fields of the tensor multiplet at the
classical level.

\vfill\eject
\pagestyle{plain}
 
%%%%%%%%%%%%%%%%%%%%%%%%%%%%%%%%%%%%%%%%%%%%%%%%%%%%%%%%%%%%%%%%%%%%%%%%%%%

\section{Introduction}
\vspace{3mm}

%%%%%%%%%%%%%%%%%%%%%%%%%%%%%%%%%%%%%%%%%%%%%%%%%%%%%%%%%%%%%%%%%%%%%%%%%%%

Globally supersymmetric field theories in various dimensions have
recently been playing new and interesting r\^{o}les in the context of
the physics of branes. There is now a large amount of literature devoted
to this subject which we shall not attempt to review here, even briefly.
It suffices to mention that such theories, or suitably nonlinearised
variations of them, are candidate worldvolume theories, at least in
principle, for superbranes. This consideration motivated Bergshoeff,
Sokatchev and one of the authors \cite{bss} to construct the coupling of
an $N=(1,0)$ tensor multiplet to Yang-Mills in six dimensions (see
\cite{bss} for a more detailed discussion of other motivations). The
construction was a direct one not employing global limits of matter
coupled $D=6$ supergravity theories known at the time \cite{ns1,ns2}. In
fact, it was not clear in \cite{bss} how to take such limits. A peculiar
aspect of the BSS model is that while the field equations of the tensor
multiplet contain the Yang-Mills field, the reverse is not the case.
Nevertheless, the model is perfectly gauge invariant and supersymmetric
at the classical level.

In a recent paper, Duff, Liu, Lu and Pope (DLLP) \cite{dllp} found a way
to take the global limit of $D=6$ supergravity coupled to Yang-Mills and
multi-tensor multiplets \cite{romans,sag1, sag2,sag3,ns3,dl} and indeed
recovered the BSS model by this means. However, DLLP also considered a
limit in which the Yang-Mills field equations do contain the tensor
multiplet fields. (See also \cite{dl} for a discussion of superspace
constraints of supergravity plus Yang-Mills system giving rise to the
BSS model in a particular limit). 

The main purpose of this paper is to show that the DLLP global limit
\cite{dllp} corresponds to the BSS model augmented with terms that
originate from the Green-Schwarz style anomaly-cancelling local
counterterms, and to exhibit the structure of what is known as the
anomaly supermultiplet involved in this case \cite{hu,bpt}. Analogous
considerations have arisen in the case of supergravity plus matter
systems in \cite{sag2} and in more detail in \cite{sag3}.

We mentioned above that a peculiar aspect of the BSS model is that while
the field equations of the tensor multiplet contain the Yang-Mills
field, the reverse is not the case. In this paper, we will show that the
BSS model has a dual formulation in which the opposite situation arises,
namely while the tensor multiplet is free, the field equations for the
Yang-Mills multiplet contain the tensor multiplet fields! This is
achieved by reversing the role of the dimensionful parameter $\a'$ that
arises in the BSS model and the Planck constant $\hbar$ that arises
when one adds the anomaly cancelling terms. The classical dual model is
perfectly gauge invariant, and the $\hbar$ dependent corrections to the
equations which make the tensor multiplet interacting arise at the
quantum level as a result of the anomaly cancellation mechanism.

After recalling the BSS model in the next section we determine the
anomaly-corrected field equations via the anomaly supermultiplet in the
subsequent sections, both in superspace and in the component formalism.
The dual formulation of the BSS model is described in the penultimate
section. Further comments on the anomaly-corrected version of the BSS
model are given in the conclusions.

%%%%%%%%%%%%%%%%%%%%%%%%%%%%%%%%%%%%%%%%%%%%%%%%%%%%%%%%%%%%%%%%%%%%%%%%%%%

\section{The Classical Tensor-Yang-Mills System}
\vspace{3mm}

%%%%%%%%%%%%%%%%%%%%%%%%%%%%%%%%%%%%%%%%%%%%%%%%%%%%%%%%%%%%%%%%%%%%%%%%%%%

Consider $(1,0)$ superspace in $D=6$ with coordinates $z^M=(x^\mu,
\theta^{\alpha i})$ where $\theta^{\alpha i}$ are symplectic
Majorana-Weyl spinors carrying the $Sp(1)$ doublet index $i=1,2$. The
basic superfields we shall consider are the supervielbein $E_M{}^A$, the
super two-form $B={1\over 2!} dz^M\wedge dz^N\, B_{NM}$ and the Lie
algebra valued Yang-Mills super one-form $A=dz^M\, A_M$. (Our
conventions for super $p$-forms are as in \cite{howe}). The torsion
super two-form $T^A$, the Yang-Mills curvature two-form $F$ and the
Chern-Simons modified super three-form ${\cal H}$ are defined as follows

\bea
&& 
T^A = dE^A\ ,\quad\quad F=dA+A\wedge A\ ,\nn\\
&&
{\cal H} = \ft12 dz^M dz^N dz^P \left(\partial_P B_{NM}-{\a'\over 2}\,
\tr (A_P F_{NM}-{2\over 3} A_P A_N A_M )\right) \ . \la{d3}
\eea

They satisfy the (flat) superspace Bianchi identities 
$ dT^A =0$, $DF=0$ and 

\be
d{\cal H} = {\a'\over 8}\, {\rm tr}\, F\wedge F \ , \la{hb}
\ee

where $\a'$ is an arbitrary dimensionful constant and $D=d+A$. In flat
$D=6$ superspace the only nonvanishing torsion component is

\be
 T_{\a i,\b j}{}^a=2\vare_{ij}(\c^a)_{\a\b}\ .
\ee

Next, we introduce the Lie algebra-valued spinor superfield $\l_{\a i}$
and a scalar superfield $\phi$. The BSS model is then characterised by
the constraints \cite{bss}

\bea
&& 
F_{\a i,\b j}=0\ ,\quad\quad 
F_{a \a i}=-(\c_a)_{\a\b} \l^{\b}_i\ ,\nn\\
&&\nn\\&& 
{\cH}_{ab\a i}=0\ ,\quad\quad {\cH}_{\a i,\b j,\c k}=0\ , \nn\\
&&\nn\\&& 
{\cH}_{a \a i,\b j}= -2 \phi\,(\c_a)_{\a\b}\,\vare_{ij}\ ,\nn\\
&&\nn\\&& 
{\cH}_{ab\a i}=-(\c_{ab})_\a{}^\b\, D_{\b i}\phi\ ,\la{cons}\\
&&\nn\\&& 
{\cH}_{abc}^{+}=(\c_{abc})^{\a\b} D_{\a}^i D_{\b i}\phi\ , \nn\\
&&\nn\\&& 
{\cH}_{abc}^{-}=\a'(\c_{abc})_{\a\b}~\tr \l^{\a i}\l^\b_i\ ,
\nn\\
&&\nn\\&& 
D_\a^{(i} D_\b^{j)} \phi=\a'\e_{\a\b\c\d}~\tr~\l^{\c(i} \l^{\d j)}\ .\nn 
\eea
where ${\cal H}_{abc}^{-}$ is anti--self--dual projected and 
${\cal H}^+_{abc}$ is  self--dual projected, i.e. ${\cal H}^\pm_{abc} 
= 1/2~({\cal H}_{abc} \pm \tilde {\cal H}_{abc}$). 
The components of the gauge spinor superfield $\l_{\a i}$
can be defined as

\be
\l^{\alpha i} = \l^{\alpha i}|\ , \quad F^{ab} = 
(\c^{ab})^\a_\b D_{\a i} \l^{\b i}|\  , 
\quad Y^{ij} = D_{\a }^{(i} \l^{\a j)}|\  , \la{ymc}
\ee

and the components of the tensor multiplet dilaton superfield $\phi$ as

\be
\s=\phi|\ ,\qquad \chi_{\a i}=
D_{\a i}\phi|\ , \qquad 
H_{abc}^{+}=\c_{abc}^{\a\b} 
D_{\a}^{i}D_{\b i} \phi|\ , \la{tc}
\ee

where $H=dB$ and the vertical bar indicates evaluation of a superfield
at $\th=0$. Using the information provided above, one then finds the
component form of the field equations \cite{bss}

\bea
{\cH}_{\m\n\r}^{-} &=& -\ft{\a'}{2} {\rm tr}\,
(\bar\l\c_{\m\n\r}\l)\ , \nn\\
[+0.15cm]
\c^\m\del_\m\chi^i &=& \a' {\rm tr}\,\left(\ft14 \c^{\m\n} F_{\m\n}
\l^i + Y^{ij}\l_j\right)\ ,\la{te}\\
[+0.15cm]
\del_\m \del^\m \s &=& \a' {\rm tr}\,\left(-\ft14 F^{\m\n}F_{\m\n} - 2\bar\l
\c^\m D_\m\l+ Y^{ij}Y_{ij}\right)\ , \nn
\eea

where $D\l =d\l+[A,\l]$. The first equation can be rewritten as 

\be
\del_\eta {\cH}^{+ \eta\l\t}= \ft{\a'}{48}\, \vare^{\l\t\m\n\r\s}\,
\tr~\left( F_{\m\n} F_{\r\s} -\ft{4}{3} \bar\l \c_{\m\n\r}
D_\s \l \right)\ .
\ee

The supersymmetry transformations of the system are
%%%
%%%%%%%%%%%%%%%%%%%%%%%%%%%%%%%%%%%%%%%%%%%%%%%%%%%%%%%%%%%%%%%%%%%
\footnote{
We use the notation and conventions of \cite{bst}. In particular,
$(A,\lambda, Y_{ij})$ take values in the Lie algebra of the
corresponding gauge group and the contraction of $Sp(1)$ indices
in fermionic bilinears is suppressed.
}
%%%%%%%%%%%%%%%%%%%%%%%%%%%%%%%%%%%%%%%%%%%%%%%%%%%%%%%%%%%%%%%%%%%%
%%%

\bea
\d A_\m&=& -\bar\e\c_\m\lambda\ ,\nn\\
[+0.15cm]
\d\l^i &=& \ft18 \c^{\m\n} F_{\m\n} \e^i - \ft12 Y^{ij}\e_j\ , \nn\\
[+0.15cm]
\d Y^{ij} &=& - {\bar\e}^{(i}\c^\m D_\m\l^{j)}\ .\la{ymt}
\eea

The corresponding rules for the on-shell self--dual tensor multiplet
coupled to Yang--Mills are 

\bea
\d\s &=& \bar\e\chi\ ,\nn\\
[+0.15cm]
\d\chi^i &=& \ft{1}{48} {\cH}^{+}_{\m\n\r} \c^{\m\n\r} \e^i 
+\ft14 (\del_\m\s) \c^\m\e^i 
- \ft{\a'}{4} {\rm tr}\,(\c^\m\l^i \bar\e\c_\m\l)\ , \nn\\
[+0.15cm]
\d B_{\m\n} &=& \a' {\rm tr}\,
A_{[\m} \d A_{\n]}-\bar\e\c_{\m\n}\chi \ , \la{tmt}
\eea

The commutator of two supersymmetry transformations \eq{tmt} closes on
all components of the tensor multiplet modulo the field equations
\eq{te}, and the supersymmetry algebra is \cite{bss}

\be
[\d(\e_1),\d(\e_2)] = \d (\xi^a)+\d (\L) + \d(\L_a)\ , \la{p1}
\ee

where the translation parameter $\xi^a$, the tensor gauge transformation
parameter $\L_a$ and the gauge parameter $\L$ are given by

\be
\xi^a=\ft12 {\bar\e_2}\c^a\e_1\ ,\qquad
\L_a=\xi^b B_{ba}+\s\xi_a\ ,\qquad \L=-\xi^a\L_a\ ,
\label{p2}
\ee
and the tensor gauge transformation takes the form
\be
\d_\L B_{\m\n}= -\ft{\a'}{2} \tr~\L( \del_\m A_\n - \del_\n A_\m)\ . \la{tgt}
\ee

The tensor multiplet is always on-shell (in fact, its off-shell version
is not known), but the Yang-Mills multiplet is still off-shell. It is
put on-shell by setting the triplet of auxiliary fields to zero,

\be
Y_{ij}=0\ , \la{y}
\ee

which then implies the Yang-Mills equations of motion

\bea
&& \c^\m D_\m \l = 0\ , \nn
\w2
&& D_\m F^{\m\n}+2[{\bar\l},\c^\n\l]=0\ . \la{yme} 
\eea

Note that these equations do not contain the fields of the tensor
multiplet. Despite this odd feature, the system of equations of
motion \eq{te} and \eq{yme} are perfectly consistent in that they do
possess classical gauge symmetry and supersymmetry. At the quantum
level, the system develops gauge anomalies due to the chiral gauge
fermions circulating in loops. For certain groups these anomalies can be
cancelled by the Green-Schwarz mechanism, and in such cases one can derive
the effect of the anomaly-cancelling counterterm as we shall see in the
next section.

%%%%%%%%%%%%%%%%%%%%%%%%%%%%%%%%%%%%%%%%%%%%%%%%%%%%%%%%%%%%%%%%%%%%%%%%%%%

\section{The Inclusion of Anomalies }

%%%%%%%%%%%%%%%%%%%%%%%%%%%%%%%%%%%%%%%%%%%%%%%%%%%%%%%%%%%%%%%%%%%%%%%%%%%

The only possible local anomaly in the model described above is the
gauge anomaly due to the minimal coupling of the Yang-Mills field with
the chiral gauge fermions. The anomaly polynomial is thus proportional
to $({\rm dim\ G})\,\tr F^4$. The associated gauge anomaly can be
cancelled by the Green-Schwarz mechanism provided that the anomaly
polynomial factorizes as $(\tr F^2)^2$. This factorization is possible
only for the gauge groups $E_8, E_7, E_6, F_4, G_2, SU(3), SU(2), U(1)$,
or any of their products with each other \cite{Okubo}. Assuming that an
anomaly free group is chosen, then the gauge transformation \eq{tgt} of the
antisymmetric tensor field can be utilized to cancel the anomaly as
usual. To this end, one would normally introduce a counterterm proportional to

\be
\hbar~B_2\wedge \tr~F\wedge F\ , \la{ct}
\ee

together with its supersymmetry partners in the one-loop effective
Lagrangian. According to \cite{sei}, one of these partner terms is

\be
\hbar~\s \tr~ F^2\ .  \la{sei}
\ee

This means that the Yang-Mills gauge coupling is quantum mechanically
modified to

\be
{1\over g^2(\s)}= {1\over g^2} + \hbar~\s\ . \la{gc}
\ee

This fact will be used later. We should note, however, that, strictly
speaking, one does not have a manifestly Lorentz invariant action (due
to the presence of the antisymmetric tensor field $B$) unless one
introduces new fields. As we do not wish to introduce new fields here,
we have to proceed in a different manner which nonetheless captures the
corrections to the equations of motion at the one-loop level. We shall
do this by considering the structure of the supercurrent multiplet for a
gauge theory in six dimensions, as this multiplet has the same structure
as the Yang-Mills equation of motion multiplet.

On quite general grounds, it has been known for sometime \cite{hu} that
the supercurrent for a six-dimensional Abelian gauge theory is a linear
superfield $J_{ij}$ defined by

\be
D_\a{}^{(i} J^{jk)}=0\ , \quad\quad J^{ij}=J^{ji}\ ,
\ee

Its components are

\be
J^{ij}=\{ j^{ij}, \eta_\a^i, V_{\m} \}\ ,
\ee

where 

\be
\del^{\m} V_{\m}=0\ .
\ee

Such a current supermultiplet couples to the prepotential for $D=6$
super-Maxwell theory, $V_{ij}=V_{ji}$. The coupling is

\be
\int d^6 x\,d^8\th\,J^{ij} V_{ij}\ .\la{jv}
\ee

Due to the constraint on the current multiplet, this interaction is
invariant under the gauge transformations

\be 
\d V_{ij}= D_\a{}^k \L_{ijk}{}^\a\ ,
\ee

where the gauge parameter superfield is totally symmetric in its $ijk$
indices. 

If there is an anomaly, the constraint on the current multiplet will be
modified to

\be
D_\a{}^{(i} J^{jk)} = A_{\a}{}^{ijk}\ . \la{ae}
\ee

The nonvanishing of $A_{\a}{}^{ijk}$ means that the component current $V_{\m}$
is no longer conserved. The anomaly superfield $A_{\a}{}^{ijk}$ defines
a possibly reducible supermultiplet which can be called the anomaly
supermultiplet \cite{hu,bpt}. Its precise form depends on the details of
the model.

The equation of motion multiplet for super-Maxwell theory has exactly
the same structure as the current multiplet as the variation of the
action will result in a coupling of the form \eq{jv}, with $V_{ij}$
replaced by its variation. In the non-Abelian case, the solution of
constraints on the superspace field strength in terms of the
prepotential $V_{ij}$ (now Lie algebra-valued) is rather complicated
\cite{k,hs't}, but, as far as the equations of motion are concerned, the
variation of Yang-Mills action will give rise to an integral of the form

\be
\int d^6x\, d^8\th\, {\rm Tr}(\d V_{ij} J^{ij})\ .
\ee

The equation of motion of multiplet, which is itself Lie algebra-valued,
now satisfies the non-Abelian conservation constraint

\be
\nab_{\a (i} J_{jk)}=0
\ee

In fact, in the classical theory, the equation of motion multiplet is
simply the superfield whose leading component is the auxiliary field
$Y_{ij}$. Note that all the components of the equation of motion
multiplet are now Lie algebra-valued, in particular, the component
$V_{\m}$ will be covariantly conserved.

To accommodate the anomaly counterterm \eq{ct} in this framework we therefore
need to amend the equation of motion multiplet by terms proportional to
$\hbar$, so that the covariant conservation condition will be modified
by an anomaly term which will now be Lie algebra-valued as well. Thus,
in the BSS model, we set 

\be
J_{ij}= (1+\hbar~\s ) Y_{ij}- 2 \hbar {\bar\l}_{(i} \chi_{j)} \ .  
\la{me}
\ee

This form of the $\hbar $ corrections is based on dimensional
considerations, as well as the chirality and $Sp(1)$ representation
content of the available spinor fields in the theory. The $\hbar \s Y$
term is motivated by the formula \eq{gc} (we have set $g=1$ for
simplicity). To see this, one should observe that the second
supervariation of \eq{me} will yield an expression which contains in
particular the term $D^\m((1+\hbar~\s)F^{\m\n}$ which has the required
form to be consistent with \eq{gc}. We shall also see that this result
is perfectly in agreement with the global limit of anomaly-free
supergravity theory in six dimensions as well \cite{dllp}. 

Next we compute the anomaly superfield $A_\a{}^{ijk}$ from \eq{ae}.
Taking the spinorial derivative of $J_{ij}$ defined in \eq{me} amounts
to its supervariation. With the aid of the supervariations provided in
Sec. 2, one finds that only the $\hbar $ dependent part of \eq{me} gives
a nonvanishing contribution (upon symmetrization in $ijk$) and the
result is simply 

\be
A_{\a ijk} = \hbar \a' ~\vare_{\a\b\c\d}~\tr\,\l^\b_{(i}\,\l^\c_j\,\l'^\d_{k)}\ , 
\la{aijk}
\ee

where $\vare_{\a\b\c\d}$ is the constant totally antisymmetric tensor,
and where we have put a prime on one of the gauge fermions to indicate that it is not involved in the trace. We have also used the Fierz identity
\be
(\c^a)_{\a\b}\,(\c_a)_{\c\d}= 2\vare_{\a\b\c\d}.
\ee

The $\l^3$ term is, therefore, the lowest dimensional component of the
anomaly supermultiplet. By acting repeatedly with the supercovariant
derivative on it, one can obtain all of its components. In particular,
one finds the covariant anomaly $ F \wedge \tr~F \wedge F $ among these
components.

The anomaly corrections to the field equations can now be obtained as
follows. To begin with, we set 

\be 
J_{ij}=0\ ,
\ee

as a consequence of which the Yang-Mills auxiliary field is no longer
vanishing but picks up $\hbar$ dependent corrections

\be
Y_{ij} = {2\hbar\over 1+\hbar \s }\,\lb_{(i}\, \chi_{j)}\ . \la{yij}
\ee

Next, we observe that the supersymmetric variation of $J_{ij}$
defined in \eq{me} takes the form

\be
\d J_{ij} = \e^{\a k}\,\left( \vare_{k(i}\,\L_{j)\a} 
+ A_{\a ijk} \right)\ , 
\la{dj}
\ee

where $A_{\a ijk} $ is the anomaly as given in \eq{aijk} and where,
suppressing the indices of the spinors, 

\bea
\L_i &= & -\left(1+\hbar\,\s\right) \c^\m D_\m \l_i 
+{1\over 24}\hbar\,{\cH}_{\m\n\r}^+ \c^{\m\n\r} \l_i 
-{1\over 2} \hbar\,(\del_\m \s) \c^\m \l_i 
+ {1\over 4} \hbar\,F_{\m\n} \c^{\m\n} \chi_i\nn
\w1
&& -{2\hbar^2\over 1+\hbar \s }\,\lb_{(i}\,\chi_{j)}\,\chi^j 
-{1\over 4} \a'\hbar\,\tr\, \left(\c^\m\l_i \lb \c_\m \l'\right)\ . \la{le}
\eea

The prime on a field means that it is not involved in a trace. The
on-shell equation \eq{me} has been used in obtaining this result. Upon
total symmetrization in $ijk$ of the terms in the parenthesis in
\eq{dj}, the first term drops out and the second term yields the anomaly
\eq{aijk}. Thus, $J_{ij}=0$ is a superfield equation modulo anomalies.
Accordingly, isolating this anomaly term in the variation of $J_{ij}$,
we can interpret 

\be
\L_{i\a}=0\ 
\ee

as the gauge fermion field equation. From \eq{le}, we thus obtain the
anomaly-corrected gauge fermion field equation:

\bea
\left(1+\hbar\,\s\right) \c^\m D_\m \l_i &=&
{1\over 24}\hbar\,{\cH}_{\m\n\r}^+ \c^{\m\n\r} \l_i 
-{1\over 2} \hbar\,(\del_\m \s) \c^\m \l_i 
+ {1\over 4} \hbar\,F_{\m\n} \c^{\m\n} \chi_i\nn
\w1
&& -{2\hbar^2\over 1+\hbar \s }\,\lb_{(i}\,\chi_{j)}\,\chi^j 
-{1\over 4} \a'\hbar\,\tr\, \left(\c^\m\l_i \lb \c_\m \l'\right)\ . \la{lef}
\eea

It is gratifying to observe that the field equation \eq{lef} agrees with
the global limit of \cite{dllp}
%%%
%%%%%%%%%%%%%%%%%%%%%%%%%%%%%%%%%%%%%%%%%%%%%%%%%%%%%%%%%%%%%%%%%%%
\footnote{
In comparing the cubic fermion terms, it is useful to note the Fierz 
identities: $\lb^{(i}\chi^{j)} \l_j=\ft34 \cb\l\l^i-\ft18 \cb\c_{\m\n}\l
\c^{\m\n}\l^i$ and $\lb^{(i}\chi^{j)} \chi_j=-\ft34 \lb\chi\chi^i+\ft18
\lb\c_{\m\n}\chi \c^{\m\n}\chi^i$.
}.
%%%%%%%%%%%%%%%%%%%%%%%%%%%%%%%%%%%%%%%%%%%%%%%%%%%%%%%%%%%%%%%%%%%%
%%%
Thus, there really are not two different global limits of the
anomaly-free 6D supergravity plus Yang-Mills system, but only one such
limit, with or without anomaly corrections proportional to $\hbar$.

There is an arbitrariness in the coefficient of the last term in \eq{lef}
\cite{sag3} which is related to the anomalies in the system. To see
this, observe that the transformation rule for the Yang-Mills auxiliary
field $Y_{ij}$ as calculated from \eq{yij} differs from the result
obtained from \eq{ymt} when one uses the anomaly modified gauge fermion
field equation \eq{lef} by $\a'\hbar\l^3$ terms. This means that the
closure of the supersymmetry algebra is spoiled by anomalies, which is
hardly surprising. 

The transformation rule for $Y_{ij}$ can be modified so that it agrees
with the transformation rule obtained from \eq{yij}, provided that
the $\a'\hbar$ dependent last term in the gauge fermion equation
\eq{lef} is removed. The required modification is

\be
\d' Y_{ij} = -{\a'\hbar \over 2(1+\hbar\s)}\, {\bar e}\c_\m\l\,\lb'_{(i}
\c^\m \l_{j)}\ .
\ee

As a consequence of this modification, the anomaly term in \eq{dj}
disappears as well but the anomaly continues to manifest itself in the 
nonclosure of the supersymmetry algebra.

Next, we calculate the Yang-Mills equation of motion from the
supersymmetry variation of \eq{lef}. Again, isolating the terms that
cannot be absorbed into the Yang-Mills equation, and therefore belong to
the anomaly supermultiplet, we can determine all the $\hbar$ corrections
to the Yang-Mills equation. The variation of the gauge fermion
equation of motion \eq{lef} takes the form

\be
\d \L_i= -\ft14\,\c^\m\e_i V_\m +{\rm anomalous\ terms} \, \la{dl}
\ee

where the anomalous terms are those which can not be absorbed into
$V_\m$ and

\bea
V^\n &=& D_\m \left[  (1+\hbar \s ) F^{\m\n} \right] 
+\ft12 \hbar\,{\cH}_{\n\r\s} F^{\r\s} + 2 (1+\hbar \s) [\lb,\c^\n \l]\nn
\w1
&& + 2\hbar D_\m \left(\cb \c^{\m\n}\l\right) 
-\ft{\a'\hbar}{2}\,\tr\, \left( \lb \c^{\n\r\s} \l \, F'_{\r\s}\right)
\eea

up to ambiguous $\a'\hbar \l^2 F$ and $\a'\hbar^2 \l^3\chi$ terms (more
on this below). Thus, setting 

\be 
V^\m=0\ , 
\ee

gives the anomaly-corrected  Yang-Mills field equation

\bea
D_\m \left[  (1+\hbar \s ) F^{\m\n} \right] &=&
-\ft12 \hbar\,{\cH}_{\n\r\s} F^{\r\s} - 2 (1+\hbar \s) [\lb,\c^\n \l]\nn
\w1
&&- 2\hbar D_\m \left(\cb \c^{\m\n}\l\right) 
+\ft{\a'\hbar}{2}\,\tr\,\left(\lb \c^{\n\r\s} \l\, F'_{\r\s}\right) \ . 
\la{ymef}
\eea

The purely bosonic part of this result agrees with the global limit of
\cite{dllp} where the bosonic part of the Yang-Mills equation is
considered. The global limit of the full supergravity coupled to tensor
plus Yang-Mills system is expected to give the fermionic terms in
\eq{ymef}. The tensor multiplet equations are given in \eq{te}, with
$Y_{ij}$ replaced by the fermionic bilinear form given in \eq{me}. The
result agrees precisely with the global limit of \cite{dllp}.

The $\a'\hbar$ dependent terms in \eq{ymef} are ambiguous due to the
possible shifting of anomalies to the supersymmetry transformation
rules. The ones exhibited in \eq{ymef} are such that the divergence of
the Yang-Mills current is purely the covariant anomaly, namely

\be
D_\m V^\m= -\ft{1}{16} \a'\hbar\, \e^{\m\n\r\s\l\t}\,
\tr\, \left(F_{\m\n}F_{\r\s} F'_{\l\t}\right) \ .
\ee

We conclude this section by commenting on the supermultiplet structure of
the anomalies. The equations of motion form an $8+8$ component linear
multiplet in the absence of anomalies. To include the effect the
anomalies, one has to enlarge the multiplet to accommodate the field
$A_{\a ijk}$ defined in \eq{aijk}. At first sight, a candidate such
multiplet is the $32+32$ component ``relaxed hypermultiplet'' constructed
sometime ago in \cite{hs't2,hst} (this is somewhat of a misnomer; it would be
more appropriate to call it the ``relaxed linear multiplet''). This
multiplet contains the $8+8$ equations of motion components $(J_{ij},
\L_{\a i}, V_\m)$ and the $24+24$ anomaly components $(L_{ijkl}, A_{\a
ijk}, V_{\m ij}, \xi_{\a i}, C)$, where the $Sp(1)$ indices are
symmetrized. The supersymmetry transformation rules for all these fields
can be found in \cite{hst}. In particular, the variation of $J_{ij}$
is seen to agree with \eq{dj}. Comparing the transformation \eq{dl} with
those of \cite{hst}, one finds that the $V_\m$ term is in agreement and
that the anomalous terms in \eq{dl} are candidate for the field $V_{\m ij}$.
However, the supersymmetric variation of $A_{\a ijk}$ must yield the
field $L_{ijkl}$ in addition to $V_{\m ij}$ according to the relaxed
hypermultiplet transformation rules given in \cite{hst}. In the model
considered here, the supersymmetry variation of the anomaly $A_{\a ijk}$
can yield the $V_{\m ij}$, but it is not clear if it can yield the field
$L_{ijkl}$. Instead a field of the form $W_{\m ijkl}$ arises and, as far
as we can see, it is not curl free. This suggests that, in order to
capture the full supermultiplet structure of the equations of motion
plus anomalies, one should relax the ``relaxed hypermultiplet'' even
further, for example by relaxing the (covariant) curl-free condition on
$W_{\m ijkl}$. This point requires further investigation.

%%%%%%%%%%%%%%%%%%%%%%%%%%%%%%%%%%%%%%%%%%%%%%%%%%%%%%%%%%%%%%%%%%%%%%%%

\section{ The Dual Model}

%%%%%%%%%%%%%%%%%%%%%%%%%%%%%%%%%%%%%%%%%%%%%%%%%%%%%%%%%%%%%%%%%%%%%%%%

We know from old results \cite{gn,ss} that the anomaly cancellation in a
dual formalism works in such a way that the anomaly polynomial is still
the same, but the r\^{o}le of the classical counterterm and $\hbar$
dependent gauge transformation get interchanged. Namely, in the dual
formulation the one loop level counterterm \eq{ct} (proportional to
$\hbar$) becomes a classical term, while the gauge variation \eq{tgt}
(which is classical in the BSS model) arises at the quantum level
(proportional to $\hbar$). 

This suggests that one makes the interchange 

\be
\a' \leftrightarrow \hbar \la{dr}
\ee

in the formulae discussed in this paper as a consequence of which the
tensor field equations now become free, while the Yang-Mills
equations have interaction terms proportional to the tensor
multiplet. In this case the Chern-Simons modification to ${\cH}$ is
proportional to $\hbar$ and consequently, the Yang-Mills equation
is anomaly free at the classical level. 

The dualization procedure just described yield the following system of
classical equations of motion. Firstly, the tensor multiplet equations of
motion become

\bea
H_{\m\n\r}^{-} &=&0\, \nn\\
[+0.15cm]
\c^\m\del_\m\chi^i &=&0\ ,\la{te2}\\
[+0.15cm]
\del_\m \del^\m \s &=& 0\ , \nn
\eea

where $H$ no longer contains the Chern-Simons form and is given by $H=dB$.
While the tensor multiplet has become free, the equations of motion for
the Yang-Mills multiplet now contain the tensor multiplet as follows

\bea
\left(1+\a'\,\s\right) \c^\m D_\m \l_i &=&
{1\over 24}\a'\,H_{\m\n\r}^+ \c^{\m\n\r} \l_i 
-{1\over 2} \a'\,(\del_\m \s) \c^\m \l_i 
+ {1\over 4} \a'\,F_{\m\n} \c^{\m\n} \chi_i\nn
\w1
&& -{2\a'^2\over 1+\a'\s }\,\lb_{(i}\,\chi_{j)}\,\chi^j \ , \la{1}
\w3
D_\m \left[  (1+\a' \s ) F^{\m\n} \right] &=&
-{1\over 2} \a'\,H^+_{\n\r\s} F^{\r\s} - 2\a' D_\m \left(\cb \c^{\m\n}
\l\right) - 2 (1+\a'\s) [\lb,\c^\n \l]\ . \la{2}
\eea

The system of equations \eq{1} and {2} are invariant under the free
tensor multiplet supersymmetry transformations

\bea
\d\s &=& \bar\e\chi\ ,\nn
\w1
\d\chi^i &=& \ft1{48} H^{+}_{\m\n\r} \c^{\m\n\r} \e^i 
+\ft14 (\del_\m\s) \c^\m\e^i \ , \nn
\w1
\d B_{\m\n} &=& -\bar\e\c_{\m\n}\chi \ ,  \la{tmt2}
\eea

and the interacting Yang-Mills multiplet supersymmetry transformations

\bea
\d A_\m&=& -\bar\e\c_\m\lambda\ ,\nn
\w1
\d\l^i &=& \ft18 \c^{\m\n} F_{\m\n} \e^i 
-{\a'\over 1+\a'\s} \lb^{(i} \chi^{j)} \e_j\ .\la{ymt2}
\eea

Note that the terms previously proportional to $\a'$ have been dropped
since they now correspond to quantum corrections proportional to $\hbar$.
The system of equations given above represent the dual formulation of
BSS model. Like the BSS model, it is perfectly consistent at the
classical level. One can check, for example, that the Yang-Mills
equation \eq{2} is consistent. To show that the divergence of the
Yang-Mills equation does not give rise to an anomaly, one uses $dH=0$
and the equations of motion for the fermions. 

\section{Comments}

The corrections to the equations of motion determined here by anomaly
considerations are only those which originate from the Green-Schwarz
anomaly cancelling {\it local} counterterms. To obtain the fully
consistent equations of motion at the quantum level one must also take
into account the non-local corrections to the one loop effective
action. This raises the question of which equations of motion are to be
solved in search of special solutions of the theory. It is clearly
necessary to check that the anomalous terms in the equations of motion,
e.g. the anomalous divergence of the Yang-Mills equation, vanish for a
given solution. However, this may not be sufficient. One should in
principle check that the remaining part of the one-loop effective action
does not spoil the solution. 

In $D=10$ where one also has a supergravity plus Yang-Mills system with
Green-Schwarz anomaly cancellation mechanism, one usually appeals to an
underlying and exactly solvable conformal field theory to argue that a
given solution may be exact. In the present case of six dimensions, one
should seek a string origin of the model. Indeed, one of the motivations
for the construction of the BSS model was that it might be a worldvolume
theory for a brane. In the last few years a great deal of progress has
been made in understanding the types of branes that can arise in
string/M-theory and in determining their worldvolume content. In
particular, it has been shown that $(1,0)$ supersymmetric matter coupled
Yang-Mills theories indeed arise in a configuration for configurations
of branes involving NS fivebranes, $D6$ and $D8$ branes. The field
content of the BSS model arises is a special case in which $D6$ branes
stretch between $NS$ fivebranes. For a summary of basic facts about this
system, see for example, \cite{bk3}.

What is still lacking is the determination of the full nonlinear
worldvolume action for the intersecting branes. That would provide an
interesting nonlinear extension of the BSS model and an appropriate
framework to study the brane within brane solitons.  

It would also be interesting to formulate the matter coupled
anomaly-free supergravity theories in six dimensions such that the
classically gauge invariant and supersymmetric part of the action (or
the equations of motion) is identified and the anomaly corrections are
determined by means of the anomaly equations. A great deal of progress
has been made in this direction in \cite{sag3}, but the task of
determining the anomaly multiplet in terms of the supergravity plus
matter multiplet fields remains to be completed. In addition to being
relevant to intersecting branes and the resulting matter plus gauge
theory dynamics, the study of the anomalies in six dimensions is also
expected to shed light on the technically far more difficult problem of
understanding the anomaly multiplet structure in ten dimensions.

\bigskip

\noindent {\bf Acknowledgements}
\vspace{.5truecm}

We thank C.S. Chu, M.J. Duff, A. Karch, S. Randjbar-Daemi, M. Rozali and
P. Sundell for stimulating discussions, and S. Ferrara for pointing out
the relation \eq{gc} to us. We also thank the Abdus Salam International
Center for theoretical Physics for hospitality.

\pagebreak

\ed